# Estimate of the number of fragments formed during the rapid radial expansion of a ring.

Goloveshkin V.A., Myagkov N.N.

*Institute of Applied Mechanics of Russian Academy of Sciences, 7 Leningradsky Prospect, 125040 Moscow, Russia; e-mail: n.myagkov@iam.ras.ru, phone +7 495 946-17-65.*

**Abstract**
In recent paper (Goloveshkin and Myagkov 2014) we proposed a two-dimensional energy-based model of fragmentation of rapidly expanding cylinder under plane strain conditions. The model allowed one to estimate the average fragment length and the number of fragments produced by ductile fracture of the cylinder. In present note we show that the proposed approach can be used to estimate the number of fragments in a problem of fragmentation of an expanding ring.

Two one-dimensional models of the dynamic fragmentation of rapidly expanding metal cylinders have been actively discussed in recent years [1]. First is the statistics-based theory of Mott and second is the energy-based theory of Grady. The models allow one to estimate both the average fragment length and the number of fragments. These models are generalized to the case of the expanding rings in a trivial way due to their one-dimensionality

In recent paper [2] we proposed a two-dimensional energy-based model of fragmentation of rapidly expanding cylinder in conditions of the ductile behavior of the material and under plane strain. The model allowed us to estimate the average fragment length and the number of fragments produced by ductile fracture of the cylinder. They obey the two-thirds power dependence on strain rate like the energy-based model of Grady. However, there is a significant difference between our model and Grady's model. The difference consists in the presence of the cylinder-wall thickness into expressions for the average fragment length and the number of fragments. In present note we show that the proposed approach [2] can be used to estimate the number of fragments in a problem of fragmentation of an expanding ring.

Let us consider the ring as a region in the cylindrical coordinate system $(r, \varphi, z)$, defined by the relations, $|z| < b$, $R - a < r < R + a$. We assume that material of the ring is incompressible with the density ρ and its mechanical behavior obeys the ideal rigid-plastic model with yield stress $Y$. When selecting the initial velocity field in the ring, we have assumed that at the initial moment the velocities are directed perpendicular to the axis of symmetry. Then, from the incompressibility condition it follows

$$u_r = \frac{V_i R}{r}, \ u_\varphi = 0, \ u_z = 0 \qquad (1)$$

where $V_i$ is a constant.

Select a fragment of the ring $|z| < b$, $R - a < r < R + a$ and $|\varphi| < \beta$, with the initial velocity field (1). Over time after the formation, the fragment will move as a rigid body with constant velocity. Using the laws of conservation of energy and momentum is easy to estimate the reduction of kinetic energy of the fragment $\Delta E = E_0 - E_\infty$ as a result of the velocity equalizing over the volume of the fragment.

$$\Delta E = \frac{4}{3}\rho V_i^2 Rba\beta^3 \qquad (2)$$

In the derivation of (2) we used the following assumptions: $a/R \ll 1$ and $\beta \ll 1$. It is seen that the expression (2) is symmetrical with respect to the size of $a$ and $b$ in the cross section of the ring.

We estimate the fracture energy $A_f$ required for the formation of a surface of discontinuity in the expansion ring. For this purpose, we use the solution of the two-dimensional model of necking considered in [2] under plane strain conditions. However, in the case under consideration, there are two options for the two-dimensional velocity field, each of them correspond to plane strain problem. The solution for each option can be easily obtained from the solution presented in [2]. As a result, the following values of $A_f$ are obtained:

$$(A_f)_1 = \frac{8}{\sqrt{3}} Y b^2 a \quad \text{и} \quad (A_f)_2 = \frac{8}{\sqrt{3}} Y b a^2, \qquad (3)$$

Obviously, the choice of the expression for the minimum value of $A_f$ depends on the aspect ratio of $a$ and $b$ in the cross-section of the ring.

In paper [2] it was shown that the potential energy of the fragment can be neglected compared to its kinetic energy. Then the energy balance due to the fragmentation has the form

$$\Delta E = (A_f)_{min} = \begin{cases} (A_f)_1, \text{if } b < a \\ (A_f)_2, \text{if } a < b \end{cases}, \qquad (4)$$

I.e., we take into account the minimum value of the fracture energy in the energy balance. Then we get from (2)-(4)

$$s = \left(16\sqrt{3}\frac{Ya}{\rho \dot{\varepsilon}_\varphi^2}\right)^{1/3} \text{ and } n = \pi R\left(\frac{\rho \dot{\varepsilon}_\varphi^2}{2\sqrt{3}\, Ya}\right)^{1/3}, \text{ if } a<b, \qquad (5)$$

and

$$s = \left(16\sqrt{3}\frac{Yb}{\rho \dot{\varepsilon}_\varphi^2}\right)^{1/3} \text{ and } n = \pi R\left(\frac{\rho \dot{\varepsilon}_\varphi^2}{2\sqrt{3}\, Yb}\right)^{1/3}, \text{ if } a>b, \qquad (6)$$

where $\dot{\varepsilon}_\varphi = V_i/R$; $s = 2R\beta$ and $n = \pi/\beta$ - the average fragment length and the number of fragments, respectively. It can be seen that the result depends on the aspect ratio of $a$ and $b$ in the cross-section of the ring. In the case of $a<b$, the estimates (5) coincide with those found for the problem of the expanding cylinder [2].

**Acknowledgements.** This research was supported by the Russian Foundation for Basic Research (project 15-01-00565).

**References**

1. D.E. Grady Fragmentation of Rings and Shells. Springer, Berlin 2006, 373 pages.
2. V.A. Goloveshkin, N.N. Myagkov. Fragmentation model for expanding cylinder. //International Journal of Fracture. 2014. V. 187. № 2. C. 239-243.